\documentstyle[aps,twocolumn,floats]{revtex}
\begin{document}
\newcommand\1{$\spadesuit$}
\newcommand\2{$\clubsuit$}
\def\be{\begin{equation}}
\def\ee{\end{equation}}
\def\ba{\begin{eqnarray}}
\def\ea{\end{eqnarray}}
\tighten
\draft
\twocolumn[\hsize\textwidth\columnwidth\hsize\csname 
@twocolumnfalse\endcsname
\title{On the Spectrum of Fluctuations in an Effective Field Theory
of the Ekpyrotic Universe}

\author{Robert Brandenberger$^{1)}$ and Fabio Finelli$^{2)}$}
\address{$^{1)}$
 Department of Physics, Brown University, Providence, RI 02912, USA.\\
e-mail: rhb@het.brown.edu}
\address{$^{2)}$
Department of Physics, Purdue University, West Lafayette, IN 47907, USA.\\
e-mail: fabio@physics.purdue.edu}

\date{August 28, 2001}
\maketitle

\begin{abstract}

We consider the four-dimensional effective field theory which 
has been used in previous studies of perturbations in
the Ekpyrotic Universe, and discuss the spectrum of
cosmological fluctuations induced on large scales by quantum fluctuations
of the bulk brane. 
By matching cosmological fluctuations on a constant energy density
hypersurface we show that the growing mode during the very slow collapsing
pre-impact phase couples
only to the decaying mode in the expanding post-impact phase,
and that hence no scale-invariant spectrum of adiabatic fluctuations is
generated. Note that our conclusions may not apply to improved
toy models for the Ekpyrotic scenario.
 
\end{abstract}

\vspace*{1cm}
]

\section{Introduction}

Recently, Khoury et al. \cite{Khoury:2001wf} suggested a new cosmological
scenario which they called the {\it Ekpyrotic Universe},
a model motivated by string theory (more precisely on
Horava-Witten theory \cite{Horava:1996qa}). The authors claim it solves
the well-known problems
of standard cosmology such as the horizon, flatness and monopole problems
(see e.g. \cite{Guth:1981zm})
without the need to invoke a period of inflationary expansion (see,
however, \cite{Kallosh:2001ai} for an opposing view concerning this
claim). In particular, a mechanism which predicts a
scale-invariant spectrum of adiabatic density fluctuations, whose magnitude
can be adjusted to agree with observations, was suggested.

According to the Ekpyrotic Universe, the Big Bang of standard cosmology
is replaced by the collision of two parallel branes, one of which is
our visible space-time, in a higher dimensional bulk. In the specific
realization studied in detail in \cite{Khoury:2001wf}, the
visible space-time is an
orbifold fixed plane in a five-dimensional bulk space-time, the fifth
dimension being a $S^1/Z_2$ orbifold whose length is substantially
larger than the radius of the internal six-dimensional Calabi-Yau
space. The second orbifold fixed plane is the hidden brane. The
scenario assumes a cold initial Universe with flat orbifold fixed
planes, and is based on the
suggestion that a light bulk brane is
nucleated near the hidden
brane, and attracted via a bulk potential to the visible brane. The
impact of the bulk brane on the visible brane generates the heating
event of the visible Universe which replaces the {\it Big Bang} of
standard cosmology \footnote{For a discussion of conceptual issues
crucial to the Ekpyrotic Universe scenario, in particular the
initial condition problem, the reader is referred to 
\cite{Kallosh:2001ai,Khoury:2001iy,Donagi:2001fs,Kallosh:2001du,Enqvist:2001}.}.

This paper concerns the
proposed mechanism of generating a scale-invariant spectrum of
adiabatic cosmological fluctuations without the need to invoke
inflation
\footnote{Note that there are toy models based on
topological defects which also
produce a scale-invariant spectrum (see e.g. 
\cite{Turok:1996wa,Avelino:2000iy}),
but these alternatives lack a convincing microphysical motivation. There
also exist some special backgrounds for which pre-big-bang cosmology
generates such a spectrum \cite{Copeland:1997ug}.}. The
suggestion is that quantum fluctuations in the position of a light bulk
brane in the five-dimensional bulk of heterotic M-theory lead to a
space-dependent time delay in the brane collision which results in
density fluctuations on the visible brane. Similar to the case of
inflationary cosmology \cite{Guth:1982ec}, small quantum fluctuations can 
generate 
a large time delay, and hence large density fluctuations. For a
bulk potential given by a single exponential factor,  
the spectrum calculated using the methods of \cite{Khoury:2001wf} (which
effectively neglect the metric fluctuations) 
is scale-invariant. 

However, Lyth has recently claimed \cite{Lyth:2001pf} that the
results change in a dramatic way when the metric fluctuations are
taken into account, and that the curvature 
fluctuations produced in the Ekpyrotic Universe are in fact not amplified
at all on super-Hubble scales and that the resulting spectrum of
curvature fluctuations is a blue spectrum with index $n \simeq 3$
instead of scale-invariant ($n = 1$). Note that a blue
spectrum is also obtained in the case of simple models 
of pre-big-bang cosmology \cite{Gasperini:1993em} which only
include the dilaton field \cite{Brustein:1995kn}. Both Khoury et al. and 
Lyth use an effective four space-time dimensional model which 
includes, in addition to the graviton, only a single scalar field (a
field which represents the location of the bulk brane in the
fifth dimension). Lyth's argument is based on the fact that there is no 
growth of curvature fluctuations on super-Hubble scales.

In this paper, we use the junction conditions for
cosmological fluctuations derived by Hwang and Vishniac
\cite{Hwang:1991an} and by Deruelle and Mukhanov
\cite{Deruelle:1995kd} (see
also \cite{Martin:1998zd}) junction
conditions for cosmological fluctuations to demonstrate that, within
the context of the effective field theory used in \cite{Khoury:2001wf} and
\cite{Lyth:2001pf}, the spectrum of curvature fluctuations at late times
is blue, and that there is no overall amplification of the amplitude 
between the times when the perturbation mode exits the Hubble radius during
the pre-impact phase and when it re-enters during the post-impact period,
the reason being that the growing mode of the pre-impact phase couples
exclusively to the decaying mode during the post-impact period. 
Note that our conclusions are based on a particular
four-dimensional effective field theory for the
Ekpyrotic scenario and may not apply to improved descriptions.

Note, however, that in the Ekpyrotic
Universe there are many matter fields which must be considered,
in addition to the scalar field representing the position of the
bulk brane. In a followup paper, we will
investigate the isocurvature modes of these fields which can be excited
by the fluctuations of the bulk brane. These modes would then, in turn,
induce a growing adiabatic component. This latter mechanism would have 
close analogies with the physics which describes  the growth of
axion fluctuations in inflationary cosmology \cite{Axenides:1983hj} (see
also \cite{Linde:1984ti}), or the
growth of axion fluctuations in pre-big-bang cosmology \cite{Copeland:1997ug}.

The outline of this paper is as follows. After a review of the 
version of the Ekpyrotic Universe scenario 
presented in \cite{Khoury:2001wf} (Section 2), we discuss (Section 3) the
effective 
field theory
considered in \cite{Khoury:2001wf} and \cite{Lyth:2001pf} and demonstrate
(Section 4),
using the matching conditions of Hwang and Vishniac \cite{Hwang:1991an}
and
Deruelle and Mukhanov \cite{Deruelle:1995kd},
the absence of an overall amplification of the spectrum of fluctuations.
We conclude with a discussion of the severe
limitations of any four-dimensional effective field theory description.

\section{Review of the Ekpyrotic Universe Scenario}

The Ekpyrotic Universe scenario is not the first attempt to combine
string theory and cosmology (see e.g. 
\cite{Easson:2000mj,Brandenberger:2001ph} for recent
overviews of various approaches to string cosmology). As will be shown below, 
it bears many
analogies, in particular in terms of the analysis of fluctuations,
with pre-big-bang cosmology \cite{Gasperini:1993em} (see
e.g. \cite{lidsey} for a recent comprehensive review). Crucial to both
scenarios is the role which the dilaton or other scalar fields play
in the cosmological scenario. In both cases, the Universe is
assumed to start in a cold, symmetric state \footnote{There is also a very 
different approach to string cosmology in which the Universe is assumed
to start out in a hot thermal state, with all degrees of freedom, including
winding strings \cite{Brandenberger:1989aj} and branes 
\cite{Alexander:2000xv}, excited. This {\it{brane gas}} 
scenario has the potential of explaining why there are only three large
spatial dimensions, but it does not (yet) provide a mechanism of
structure formation.}, although in the case of
the pre-big-bang scenario this assumption can be somewhat relaxed 
\cite{Buonanno:1999bi}.
Since in this case the Universe has infinite age, the horizon problem
is solved. The assumption of initial symmetry produces
a Universe without spatial curvature \footnote{However, this
alone does not explain a second aspect of the ``flatness
problem'' which is naturally solved by inflation, namely why the 
spatial size of the visible brane
is large enough to agree with present observations, or, equivalently,
why the entropy is so large \cite{Kallosh:2001ai}.} 

The basic setup is the following. We start from the action of 11-d supergravity
compactified on a $S^1/Z_2$ orbifold and compactify the theory on a
Calabi-Yau three-fold \cite{Horava:1996qa}. The radius of the orbifold is
taken to
be much larger than the radius of the Calabi-Yau three-fold, implying that
there is a large energy range where an effective 5-d description of the
physics applies \cite{Lukas:1998fg}. The bulk effective action takes the form
\begin{equation} \label{action}
S \, = \, {{M_5} \over 2} \int_{{\cal{M}}_5} d^5x \sqrt{-g}
\bigl(R - {1 \over 2} (\partial \phi)^2 - {3 \over 2} {{e^{2 \phi}
{\cal{F}}^2} \over {5!}} \bigr) \, , 
\end{equation}
where the scalar modulus field $\phi$ represents the radius of the Calabi-Yau
three-fold, $\cal{F}$ is the field strength of a four-form gauge field, and 
the rest of the notation is standard. Note that the dilaton is
assumed to be frozen. 
 
There are two orbifold fixed planes,
the visible brane located at $y = 0$, and the hidden brane located at $y = R$,
where $y$ is the coordinate in the orbifold direction, and $R$ is the
orbifold radius. Associated with each orbifold fixed plane is a brane
action, whose form will not be important in the following. Important for
the following discussion is the fact that several fields are present in
the bulk, setting the stage for the possible generation of entropy
fluctuations.

The above action admits a symmetric BPS solution which contains, in
addition to the orbifold fixed planes, a bulk brane
parallel to the boundary fixed planes, located at the position $y = Y$. 
The solution takes the form \cite{Khoury:2001wf,Kallosh:2001ai} 
(generalized from \cite{Lukas:1998fg})
\begin{eqnarray} \label{BPSsol}
ds^2 \, &=& \, D(y)\bigl( -N^2 d\tau^2 + A^2 d{\bf x}^2) + B^2 D^4(y) dy^2
\bigr) \nonumber \\
e^{\phi} \, &=& \, B D^3(y) \\
{\cal F}_{0123y} \, &=& \,  
- A^3 N B^{-1} \alpha D^{-2}(y) \,\,\, (y < Y) \nonumber \\
&=& \, - A^3 N B^{-1} (\alpha - \beta) D^{-2}(y) \,\,\, (y > Y) \,
,\nonumber
\end{eqnarray}
where $\tau$ is conformal time and
\begin{eqnarray} \label{BPSsol2}
D(y) \, &=& \, \alpha y + C  \,\,\, (y < Y)\\
&=& \, (\alpha - \beta) y + C + \beta Y \,\,\, (y > Y) \, ,\nonumber
\end{eqnarray}
with $A, B, C, N$ and $Y$ being constants and $C > 0$. The associated
tensions of the visible, hidden and bulk branes are $- \alpha M_5^3$, 
$(\alpha - \beta)M_5^3$ and $\beta M_5^3$, respectively, where both
$\alpha$ and $\beta$ are positive.

The action (\ref{action}) does not include non-perturbative interactions
between the bulk and boundary branes. Such are assumed 
to lead to a potential $V(Y)$ with which the bulk brane is
attracted to the visible brane. If the bulk brane is
light and the interaction forces are weak, then the dynamics of a
configuration starting out in the BPS state (\ref{BPSsol}) corresponds
to the bulk brane moving adiabatically towards the visible brane. This
dynamics can be described \cite{Khoury:2001wf} in terms of a moduli space 
action
in which the parameters $A, B, C, N$ and $Y$ of the ansatz (\ref{BPSsol})
are taken to be time dependent. The effective action for these parameters
is obtained by inserting the ansatz (\ref{BPSsol}) into the action 
(\ref{action}) and integrating over space. If ${\bf x}$ dependence
of the parameters is to be allowed (which we must when studying fluctuations),
then only integration over the orbifold direction $y$ is performed.
In either case, the potential $V(Y)$ must be added to the effective
Lagrangian by hand. In \cite{Khoury:2001wf}, the parameters $B$ and $C$ 
were left
to be constant, yielding an 4-d effective action for a single ``scalar field''
$Y$ in a cosmological background with scale factor $a$ determined by 
$A$ and $N$.

In the Ekpyrotic Universe scenario,
branes and bulk are assumed to start out in the cold, symmetric BPS state
given by (\ref{BPSsol}).
It is assumed that the light bulk brane is
nucleated near the hidden brane, and that it then slowly moves
towards the visible brane, attracted by the potential $V(Y)$.
 Associated with the bulk brane is
another brane action. After integrating over the orbifold direction,
this action yields a four-dimensional effective action for a scalar field
$\varphi$ related to the orbifold position $Y$. The potential
produced by nonperturbative effects in string theory was taken to
be (see e.g. \cite{Moore:2000fs} for some work on the origin of such
potentials)
\begin{equation} \label{pot}
V(\varphi) \, = \, - V_0 e^{- \sqrt{2 \over p}(m_p)^{-1} \varphi} \, ,
\end{equation}
where $0 < p \ll 1$ and $m_p$ denotes the 4-d Planck mass  
(using the notation of \cite{Lyth:2001pf}).

The proposed mechanism of structure formation is based on quantum
vacuum fluctuations in the position of the bulk brane \footnote{The initial classical fluctuations must be tuned to be smaller than the induced
quantum vacuum perturbations \cite{Kallosh:2001ai}.}. In the 4-d
effective field theory description, these fluctuations are
represented as perturbations of the scalar field $\varphi$.
The analysis of fluctuations performed in 
\cite{Khoury:2001wf,Kallosh:2001ai,Lyth:2001pf} was based
on considering the 4-d effective action for gravity in the presence of
the scalar field $\varphi$. In \cite{Khoury:2001wf,Kallosh:2001ai}, the time 
delay
formalism of \cite{Guth:1982ec} was used to determine the fluctuations in
the time of impact of the bulk brane on the visible brane, which in turn
determines the resulting density fluctuations. The result is a 
scale-invariant spectrum of adiabatic fluctuations. The time delay method, 
however, is not accurate when the expansion of space is not
quasi-exponential \cite{Wang:1997cw}, as in this case and
similar to power law inflation. Lyth
\cite{Lyth:2001pf}, on the
other hand, focused on $\zeta$, the curvature perturbation in
comoving coordinates \cite{Bardeen:1980kt} (denoted by $-{\cal{R}}$ in
\cite{Lyth:2001pf} - we follow the notation of \cite{Mukhanov:1992me}),
a quantity which is conserved on scales larger than the Hubble
radius in the absence of non-adiabatic perturbations 
\cite{Bardeen:1980kt,Bardeen:1983qw,Brandenberger:1984tg,Lyth:1985gv}. He
showed that because $\zeta$ is conserved the curvature
fluctuations remain small in amplitude and have a blue spectrum, as
they do in the pre-big-bang models which only take into
account the dilaton field (Lyth's calculations confirm that
in the absence of metric fluctuations the power spectrum of fluctuations
in $\varphi$ is scale-invariant).

However, from the form of the BPS solution (\ref{BPSsol}) it is clear
that fluctuations in the position of the bulk brane do not decouple
from fluctuations in other fields such as the modulus field
$\Phi$. Including the
dynamics of these fields may give a different dynamics on which
the fluctuations evolve. Neglecting
these coupling means eliminating the isocurvature modes which are
inevitably present in the Ekpyrotic Universe and which will be
discussed in a followup paper. These latter allow for
the establishment and growth of entropy fluctuations on super-Hubble
scales, in close analogy to what occurs in the case of axion fluctuations
in inflationary cosmology (see e.g. \cite{Axenides:1983hj}) and in the case 
of the
growth of fluctuations on super-Hubble scales during the reheating period of 
inflationary cosmology \cite{Bassett:1999wg,Finelli:1999bu} 
(such growth does not occur in 
single field models \cite{Finelli:1999bu,Parry:1999pn}, but it does occur 
in the case of the
existence of unsuppressed entropic fluctuations 
\cite{Bassett:2000cg,Finelli:2000ya}).

\section{The Original Effective Field Theory for the Ekpyrotic Universe}

The effective action for fluctuations used in \cite{Khoury:2001wf}
and \cite{Lyth:2001pf} is
\begin{equation} \label{effaction1}
S \, = \,  \int d^4x \sqrt{-g} 
\bigl({1 \over {2 \kappa^2}} R 
 + {1 \over 2}(\partial \varphi)^2 - V(\varphi) \bigr) \, , 
\end{equation}
where, as mentioned before, $\varphi$ is a scalar field which
represents the position of the bulk brane in the fifth dimension,
and where $\kappa^2 =  m_p^{-2} = 8 \pi G$.In this effective field
theory, fluctuations in the position of the bulk brane generate
adiabatic curvature perturbations.

The background equations of motion relevant to the following analysis
are
\begin{equation} \label{bgeom1}
{\ddot{\varphi}} + 3 H {\dot{\varphi}} + V^{'} \, = \, 0 \, ,
\end{equation}
and
\begin{equation} \label{bgeom2}
H^2 \, = \, {1 \over {3 m_p^2}} \bigl( {1 \over 2} {\dot{\varphi}}^2 + 
V(\varphi) \bigr) \, ,
\end{equation}
where an overdot denotes the derivative with respect to physical time,
the prime in $V^{'}$ denotes the derivative with respect to $\varphi$,
and $H$ is the effective Hubble constant.

The relevant solution of (\ref{bgeom1},\ref{bgeom2}) is given by
\begin{equation}
a(t) \, \propto \, (-t)^p 
\end{equation} 
or
\begin{equation} \label{bgsol}
H \, = \, {p \over t} \, .
\end{equation}
In this solution, the time $t$ is negative. Since
\begin{equation}
m_p^2 {\dot{H}} \, = \, - {1 \over 2} {\dot{\varphi}}^2 \, ,
\end{equation}
we can determine the solution for ${\dot{\varphi}}$
\begin{equation} \label{phider}
{\dot{\varphi}}^2 \, = \, {{2 p m_p^2} \over {t^2}} \, ,
\end{equation}
from which the background values of all relevant quantities can be
determined. In particular,
\begin{eqnarray} 
V \, &=& \, - (1 - 3p)p{{m_p^2} \over {t^2}} \label{poteval} \\
V^{'} \, &=& \, \sqrt{2p} (1 - 3p){{m_p} \over {t^2}} \, \label{potder} .
\end{eqnarray}

In order to determine the evolution of the curvature fluctuations, we
make use of the gauge-invariant formalism of \cite{Mukhanov:1992me}
and focus on the gravitational potential $\Phi$ which appears in
the metric which includes linear fluctuations (in longitudinal gauge) as
\begin{equation} \label{longitudinal}
ds^2 \, = \, a^2(\eta)\bigl[(1 + 2\Phi)d\eta^2 - (1 - 2\Phi)dx^idx_i \bigr]
\, ,
\end{equation}
where $\eta$ is conformal time. By inserting this metric and the
decomposition of the scalar field $\varphi$ into background and
fluctuation $\delta \varphi$, it was shown by Mukhanov \cite{Mukhanov:1988jd}
(see also \cite{Sasaki:1986hm}) 
that the following combination $v$ of metric and
matter fluctuations corresponds to the canonically normalized
fluctuation variable
\begin{equation}
v \, =  \, a \bigl( \delta \varphi + {{{\dot \varphi}} \over H} \Phi
\bigr) \equiv a Q \, . 
\end{equation}

In the absence of entropy fluctuations (e.g. in the single matter
field model considered in this section), the equations of motion of
fluctuations reduce to
\begin{eqnarray} \label{basiceom}
v^{''}_k + \bigl(&k^2& - {{a^{''}} \over a} + a^2 V^{''} \\
&+& 2 a^2 \bigl[{{{\dot H}} \over H} + 3H \bigr]^{\cdot} \bigr) v_k = 0
\, , \nonumber
\end{eqnarray}
where primes on functions of time denote the derivative with respect to
conformal time, and the primes on functions of $\varphi$ denote 
derivatives with respect to $\varphi$. Overdots denote derivatives with
respect to physical time.

From the background equations of motion it follows that the last
two terms on the left hand side of (\ref{basiceom}) cancel out.
The only remaining term in the frequency for $v$ is therefore the curvature,
i.e. $a''/a$:
\be
\frac{a''}{a} = p \frac{2p-1}{(p-1)^2 \eta^2} \, .
\label{asecond}
\ee
The solution for $Q$ normalized by the initial vacuum state condition 
can thus be expressed in terms of
Hankel functions $H_\nu$:
\be
Q_k = A (-\eta)^\nu H_\nu^{(1)} (-k\eta)
\ee 
where the index $\nu$ and amplitude $A$ are given by
\be
\nu = \frac{1}{2} \frac{1-3p}{1-p}\, , \quad A =
e^{i(\nu+\frac{1}{2})\frac{\pi}{2}} \frac{\sqrt{\pi}}{2} 
\frac{1}{(1-p)^\frac{p}{1-p}} \, .
\label{nuindex}
\ee
The above solution agrees with the Minkowski limit for $- k \eta
\rightarrow \infty$. For long wavelengths ($- k \eta \rightarrow 0$), 
$Q$ scales as \cite{abrasteg}:
\be
Q_k \simeq A (-\eta)^\nu \left[ 
\frac{(-k\eta/2)^\nu}{\Gamma(\nu + 1)} - i \frac{\Gamma(\nu)}{\pi
(-k\eta/2)^\nu} \right] \, .
\label{longw}
\ee

The curvature perturbation $\zeta$ introduced by Bardeen
\cite{Bardeen:1980kt} is related to $Q$ as follows:
\be
\zeta = \frac{H}{\dot \phi} Q \, .
\ee
By using the behavior for long wavelengths (\ref{longw})
and inserting the background values for $H$ and ${\dot \varphi}$,
one obtains for the spectrum of curvature perturbations:
\begin{eqnarray}
P_{\zeta}(k) \, &=& \, {{k^3} \over {2 \pi^2}} |\zeta_k|^2 \\
&=& \frac{p \, k^{3 - 2\nu}}{4 m_p^2 \pi^4} |A|^2 \, 2^{2\nu}
\, \Gamma(\nu)^2 \, . \nonumber
\end{eqnarray}
Note that the spectrum is independent of time, a 
well-known result 
\cite{Bardeen:1983qw,Brandenberger:1984tg,Lyth:1985gv}, and that
its amplitude is proportional to $p$.

For extremely slow contraction \cite{Khoury:2001wf}, i.e. 
$p \sim 0$ or equivalently $\nu \sim 1/2$, one
obtains a blue spectrum of curvature perturbations with index $n \sim
3$ and with an amplitude which is suppressed by the small number $p$, 
in agreement with the result of Lyth. 
The reason for this is the crucial effect due to metric
perturbations in Eq. (\ref{basiceom}). The bulk brane fluctuations in
rigid spacetime, i.e. without the inclusion of metric perturbations,
are amplified by the negative effective mass
\cite{Khoury:2001wf,Kallosh:2001ai} and obtain a scale invariant spectrum.
However, when one include self-consistently the metric perturbations, the
instability due to the negative effective mass is exactly cancelled by the 
gravitational contribution \footnote{A similar exact cancellation
between the potential term and the gravitational corrections also
occurs in power-law inflation \cite{stewlyth:1992}.}, 
and the bulk brane fluctuations satisfy the same
equation of a massless minimally coupled scalar field. If contraction is
very slow, the curvature term (\ref{asecond}) in Eq. (\ref{basiceom}) is
very inefficient in amplifying fluctuations and $v$ (or $Q$) essentially 
remains in its initial vacuum state. In principle a scale invariant
spectrum for
$Q$ can be obtained for $p=2/3$ (not constant in time) or for $p >> 1$
(constant in time). In the latter case it is not clear if such 
a fast contraction can successfully solve the horizon problem.

On the other hand (as pointed out in \cite{Kost4}), the 
generalized Newtonian (Bardeen) potential $\Phi$ does grow during
the Ekpyrotic phase. The general solution for $\Phi$ on scales
much larger than the Hubble radius is \cite{Mukhanov:1992me}
\begin{equation} \label{gensol}
\Phi \, = \, S {H \over a} + D \bigl({1 \over a} \int{a dt} \bigr)^{\cdot}
\, ,   
\end{equation}
where $S$ and $D$ are constants. In an expanding Universe, e.g. in
inflationary cosmology, $S$ is the coefficient of the decaying mode,
and $D$ is the coefficient of the dominant mode. In a
contracting background, e.g. in the pre-impact phase of the
Ekpyrotic Universe or in the pre-bounce phase of pre-big-bang
cosmology, the S-mode is a growing mode. For a scale factor
described by a power of time, the D-mode is constant. This is the
case both in an expanding and in a contracting background.

In our case the scale factor is approximately constant (it
decays with a very small power of $\eta$) the growing mode
scales as $\eta^{-1}$. The growth starts at the time $\eta_k$ when the
scale labelled by $k$ exits the Hubble radius 
\begin{equation}
\eta_k \, \propto \, {1 \over k} \, .
\end{equation}
Hence, the spectrum of the growing mode of $\Phi$ is scale-invariant
\begin{eqnarray}
P_{\Phi}(k, \eta) \, &=& \, {{k^3} \over {2 \pi^2}} |\Phi_k(\eta)|^2
\\
&=& {{k^3} \over {2 \pi^2}} \bigl({{\eta_k} \over {\eta}}\bigr)^2
|\Phi_k(\eta_k)|^2 \nonumber \\
&\propto& k^0 \, . \nonumber
\end{eqnarray}
Note that the spectrum of the D-mode of $\Phi$ does not obtain
the factor $\eta_k^2$ and hence is blue ($n = 3$).

The results of the two previous paragraphs can be obtained from our 
earlier equations concerning $\zeta$ by 
relating the generalized Newtonian potential to the time variation of
$\zeta$ in the usual way \cite{Finelli:1999bu}:
\be \label{Phieq}
k^2 \Phi_k = - \frac{1}{2 m_p^2} \frac{\varphi^{' 2}}{\cal H} \zeta'_k
\,.
\ee
By using 
\be
\zeta'_k = \sqrt{\frac{p}{2}} \frac{k A}{m_p} (-\eta)^\nu 
H_{\nu-1}^{(1)} (-k\eta)
\ee
and Eq. (\ref{Phieq}) 
one obtains for the Newtonian potential for small $k$:
\be
\Phi_k = - \frac{A \sqrt{2 p}}{m_p (1-p)} 
\left[ \frac{(-\eta)^{2 \nu - 2}}{2^\nu \Gamma(\nu) k^{2 -\nu}} - i
\frac{2^{\nu -2} \Gamma(\nu-1)}{\pi k^\nu}  
\right] \,.
\ee
From this equation it is clear that $\Phi$ obtains a component
constant in time $\sim k^{-\nu}$ and a growing mode with spectrum 
$\sim k^{\nu - 2}$. This latter component for $p \sim 0$ gives rise
to a scale-invariant spectrum. 

The natural question which we are going to address in the next section 
is the fate of this growing mode. Of course, such growing modes in
contracting universes are {\em physical}. The gauge is completely
specified in Eq. (\ref{longitudinal}) and thus $\Phi$ for nonzero $k$ 
has a gauge-invariant meaning. Since $\Phi$ - and not only $\zeta$ - 
enters in observables such as the CMB anisotropies without any
suppression factors, its spectrum leaves a direct
imprint. Therefore, the only way that the growing mode of $\Phi$ in
the contracting phase can disappear from physical observables at late
times is if this mode 
only matches to the decaying mode of the post-impact radiation-dominated 
era. We will show that this is exactly what occurs.

A similar effect arises in the pre-big-bang scenario. 
Also in that context, the growing mode of the contracting phase of
the pre-big-bang phase
does not match to the constant mode of the standard big-bang phase 
\cite{Deruelle:1995kd} (It is not an issue of ``gauging away'' the
growing mode as stated in \cite{Brustein:1995kn}).

\section{Matching Conditions}

As we have seen in the previous section, the 
growing mode of the Bardeen potential $\Phi$
is increasing during the Ekpyrotic phase and obtains a
scale-invariant spectrum, whereas the curvature parameter $\zeta$
does not grow and is not scale-invariant (it has $n = 3$). This
situation is similar to what occurs in pre-big-bang cosmology
\cite{Brustein:1995kn}. During the radiation dominated
post-impact phase, both $\zeta$ and the non-decaying mode of $\Phi$ are
constant. Lyth \cite{Lyth:2001pf} argues that $\zeta$ is continuous
across the transition and that hence the final curvature
fluctuations are not scale invariant and have negligible amplitude.
However, since the variable $\Phi$ enters directly in the
expression for the cosmic microwave anisotropies, one could argue
that the scale-invariance of the growing mode of $\Phi$ will lead
to scale-invariance of late time cosmological observables (this is
the claim made in \cite{Kost4}).

In the following we will use the Hwang-Vishniac
\cite{Hwang:1991an} and Deruelle-Mukhanov \cite{Deruelle:1995kd} 
matching conditions 
for cosmological fluctuations to demonstrate that
the pre-impact growing mode of $\Phi$ does not couple to the
post-impact dominant (i.e. non-decaying) mode of $\Phi$. Since
the subdominant (constant) mode of $\Phi$ during the pre-impact
phase is not scale-invariant, this implies that the 
spectrum of $\Phi$ at late times is determined by the
initial spectrum of the constant mode of $\Phi$, and that
it is hence neither large in amplitude nor scale-invariant,
thus confirming the results of Lyth.

The Deruelle-Mukhanov matching conditions which relate the
fluctuations across a discontinuous transition in the
equation of state are applicable if the transition occurs
on a surface of constant energy density. This is satisfied
in the pre-big-bang scenario since (presumably) the graceful
exit mechanism which determines the bounce occurs at some
fixed energy density. In the Ekpyrotic scenario, the transition
surface is determined by when the bulk brane impacts on the visible
brane. Since the energy of the visible brane before impact is
constant in space and time (given by the brane tension) and since
for long-wavelength fluctuations the extra energy density imparted
to the visible brane by the collision with the bulk brane is independent
of space when measured at the time of impact, the above condition for
applicability of the Deruelle-Mukhanov matching conditions is
satisfied. The matching conditions imply that both $\Phi$ and 
\begin{equation}
{\hat{\zeta}} =
\zeta - {1 \over 3} k^2 \Phi \bigl({\cal H}^{'} - {\cal H}^2\bigr)^{-1}
\end{equation}
are continuous across the transition. For large-scale
fluctuations, this implies the continuity of $\Phi$ and $\zeta$.
Since the total
energy density does not change at the impact time, we take the magnitude of
${\cal H}$ to be continuous across the transition \footnote{Note
that we are matching a contracting phase of cosmological evolution
to an expanding phase at the impact time, similar to what is done
in the Einstein frame description of Pre-Big-Bang cosmology. The
impact in the model of \cite{Khoury:2001wf} occurs at a surface of
finite density.}
 
We write the generalized Newtonian potential $\Phi$ before impact
as
\begin{equation}
\Phi_{-} = D_{-} + S_{-} \frac{\cal{H}_{-}}{a^2}
\end{equation}
and after impact as
\begin{equation}
\Phi_{+} = D_{+} + S_{+} \frac{\cal{H}_{+}}{a^2} \,.
\end{equation}
Here, the coefficients $D_i$ ($i = \pm$) denote the amplitudes of the constant
modes, $S_{-}$ is the amplitude of the growing mode during the
Ekpyrotic phase, and $S_{+}$ is that of the decaying mode after
the Ekpyrotic phase. Subscripts $-$ and $+$ denote the limiting values of
the quantities evaluated at the impact time on the pre-impact and 
post-impact branches. 
 
We find $D_{+}, S_{+}$ as a function of $D_{-}, S_{-}$ by matching $\Phi$
and $\hat{\zeta}$.It is useful to recall the expression for $\zeta$ in
terms of $\Phi$ \cite{Mukhanov:1992me}:
\begin{equation}
\zeta \, = \, {2 \over 3} {{\Phi + H^{-1}{\dot \Phi}} \over {1 + w}}
+ \Phi \, .
\end{equation} 
Setting $k = 0$ leads to the following system of equations:
\begin{eqnarray}
& & D_{+} = D_{-} + \frac{{\cal H}_{-} S_{-} - {\cal H}_{+} S_{+}}{a^2}
\label{uno}
\\
& &\frac{\cal{H}_{-}}{{\cal H}'_{-} - {\cal H}^2_{-}}
\left( {\cal H}_{-} D_{-} + \frac{S_{-}}{a^2} ({\cal H}'_{-} - 
{\cal H}^2_{-}) \right) \nonumber \\
 &=& \frac{\cal{H}_{+}}{{\cal H}'_{+} - {\cal H}^2_{+}}
\left ( {\cal H}_{+} D_{+} + \frac{S_{+}}{a^2} ({\cal H}'_{+} - 
{\cal H}^2_{+}) \right)
\end{eqnarray}

The second equation can be rewritten as:
\begin{equation}
\frac{{\cal H}_{+}^2}{{\cal H}'_{+} - {\cal H}^2_{+}} D_{+}
= \frac{{\cal H}_{-}^2}{{\cal H}'_{-} - {\cal H}^2_{-}} D_{-}  +
\frac{S_{-} {\cal H}_{-} - S_{+} {\cal H}_{+}}{a^2} \, .
\label{due}
\end{equation}

We subtract Eq. (\ref{uno}) from Eq. (\ref{due}) and we get:
\begin{equation}
\frac{{\cal H}'_{+} - 2{\cal H}^2_{+}}{{\cal H}'_{+} - {\cal H}^2_{+}}
D_{+} =
\frac{{\cal H}'_{-} - 2 {\cal H}^2_{-}}{{\cal H}'_{-} - {\cal H}^2_{-}}
D_{-} \, . \label{tre}
\end{equation}
Thus, the late-time constant mode receives no contribution from
the pre-impact increasing mode. Hence, the final amplitude of
$\Phi$ is small, and its spectrum is not scale-invariant. This
result agrees with the conclusions of Deruelle and Mukhanov in the
case of the pre-big-bang theory. Note that in the case of
an inflationary Universe modeled with an
equation of state with $w$ constant during and after
inflation, but with a discontinuous change in
the equation of state at the end of the period of inflation, then 
Eq. (\ref{tre}) reproduces the well-known result 
\cite{Bardeen:1983qw,Brandenberger:1984tg} that the value of $\Phi$
increases by the ratio of the factors $1 + w$ between the
period of inflation and the radiation-dominated phase.

If we now take into account the term in $\zeta$ proportional 
to $k^2$, then  Eq. (\ref{due}) becomes
\begin{eqnarray}
\frac{1}{{\cal H}'_{-} - {\cal H}^2_{-}}
\left( ({\cal H}^2_{-} + \frac{k^2}{3}) D_{-} +
\frac{{\cal H}_{-} S_{-}}{a^2} 
({\cal H}'_{-} - {\cal H}^2_{-} + \frac{k^2}{3}) \right) = \nonumber \\
\frac{1}{{\cal
H}'_{+} - {\cal H}^2_{+}}
\left ( ({\cal H}_{+} + \frac{k^2}{3}) D_{+} + 
\frac{\cal{H}_{+} S_{+}}{a^2} 
({\cal H}'_{+} - {\cal H}^2_{+} + \frac{k^2}{3}) \right) \, . \nonumber
\end{eqnarray}
This change leads to
\begin{equation}
D_{+} = A D_{-} + B k^2 (S_{+} - S_{-})
\end{equation}
which means that the imprint of the growing mode during the contracting
phase on the constant mode of the standard big bang phase is suppressed by
$k^2$ and is therefore unimportant in the long wavelength limit for 
$p \sim 0$. However, we note that for $p \sim 2/3$ (for this
case $\zeta$ has a scale invariant spectrum) there is the possibility
to get a scale invariant spectrum for $\Phi$ in the expanding phase by
the matching with the growing mode in the contracting phase
(the growing mode is $\Phi \sim (k^{7/2} \eta^5)^{-1}$).

\section{Discussion}

Perhaps the most important aspect of the proposed Ekpyrotic
Universe scenario is the claim that it gives rise to a mechanism
to produce a scale-invariant spectrum of adiabatic cosmological
fluctuations with an amplitude which can be tuned to agree with
observations. 

We have studied a four-dimensional toy model of the Ekpyrotic scenario. In
this context, we have used the Hwang-Vishniac \cite{Hwang:1991an} and
Deruelle-Mukhanov
\cite{Deruelle:1995kd} matching
conditions to demonstrate that there is indeed a growing mode
of the adiabatic metric fluctuations (quantified most
conveniently in terms of the generalized Newtonian potential
$\Phi$ during the pre-impact phase, and that this mode has
a scale-invariant spectrum
(in agreement with the conclusions of \cite{Kost4}), but that
this mode does not couple to the non-decaying mode of the
post-impact phase. The non-growing mode of $\Phi$ during the
pre-impact phase does have a blue spectrum ($n = 3$). The 
final spectrum of fluctuations in the radiation-dominated phase
is thus blue and of negligible amplitude. These conclusions
can also be obtained by focusing on the variable $\zeta$. Our
work confirms that $\zeta$ is indeed constant across the transition,
(as argued in \cite{Lyth:2001pf,Lyth2}). We warn the reader that our
conclusions are based on the toy model studied, and may change
in a calculation which better encodes the
higher-dimensional stringy aspects of the Ekpyrotic scenario.

Note that our approach to this four-dimensional toy model of the Ekpyrotic
Universe assumes that
the matching happens at a finite value of the effective
four-dimensional scale factor, as suggested in \cite{Khoury:2001wf}.
It is also based on using the negative potential at the impact point,
whereas in \cite{Khoury:2001wf} it is
suggested that the potential should vanish close to $Y = 0$, as
has been emphasized in \cite{Kost4}. However, this modification of
the potential alone is unlikely to yield a scale-invariant
spectrum. It makes the model look more like the Pre-Big-Bang
scenario, and also yields a blue spectrum.

Our matching conditions applied to a toy model of inflation with a
jump of the equation of state at reheating (as opposed to a
continuous change), but with constant equation of state before
and after reheating, give the correct result
\cite{Hwang:1991an,Deruelle:1995kd}, whereas the matching proposed
in \cite{Kost4}, namely the continuous matching of the two
modes of the fluctuation variable $\epsilon_m$ (the energy density
contrast on comoving hypersurfaces, which is directly
related to $\Phi$ by a function of the background scale factor)
does not. In a new
version of the Ekpyrotic Universe \cite{Kost5} published after
this manuscript was completed, it is proposed that the big bang
of the visible Universe results from the collision of the two
boundary branes, at which point the four-dimensional scale
factor vanishes. In this case any
four-dimensional toy model will break down completely and matching
fluctuations at this singular point seems hazardous.

Another caveat is that this four-dimensional effective theory
is (as emphasized in \cite{Kallosh:2001du}) 
not consistent since it involves a sudden transition between
contraction and expansion which is not consistent with the
background Einstein equations. However, we think that by adopting the
matching conditions chosen here, spurious effects due to
this sudden transition are avoided.
Furthermore, there are
many effects concerning fluctuations in the 
five-dimensional bulk which cannot be captured in a
four-dimensional effective theory of perturbations. The
final word concerning the spectrum of fluctuations in the Ekpyrotic
Universe scenario will have to come from a full 5-d analysis
of the fluctuation equations, using e.g. the formalism recently
proposed in 
\cite{vandeBruck:2000ju,Maartens:2000fg,Langlois:2000ia,Kodama:2000fa}. 
These calculations, in turn, will have to
be based on a consistent 5-d analysis of the background dynamics.

\vspace{0.5cm} 
\centerline{\bf Acknowledgments}
\vspace{0.2cm}

This research was supported in 
part by the U.S. Department of Energy under Contracts 
DE-FG02-91ER40688, TASK A (Brown Univ.) and DE-FG02-91ER40681, 
TASK B (Purdue Univ.). Both authors thank Salman Habib for
hospitality at the 2001 Summer Workshop on Cosmology during which
some of this work was done. One of us (F.F.) is grateful to
Raul Abramo for useful comments.
R.B. thanks
David Lyth and Paul Steinhardt for communicating to him
early drafts of their unpublished papers on this subject,
and Andrei Linde and David Lyth for important feedback
on an early set of notes. He is grateful to
Nathalie Deruelle, Jai-chan Hwang, David Wands and
in particular Neil Turok for fruitful discussions on the
issues presented here, and Lev Kofman and Burt Ovrut
for feedback on the manuscript. He thanks the
organizers of the M-theory cosmology conference in
Cambridge for the opportunity to debate these
issues, 
the Department of Physics and Astronomy of the 
University of British Columbia, and in particular Bill Unruh, for
hospitality during the time this paper was written, and the
UBC theory postdocs Emil Akhmedov, Sumati Surya and Kostya
Zarembo for encouragement.




\begin{thebibliography}{}


\bibitem{Khoury:2001wf}
J.~Khoury, B.~A.~Ovrut, P.~J.~Steinhardt and N.~Turok,
``The ekpyrotic universe: Colliding branes and the origin of the hot big  bang,''
hep-th/0103239.


\bibitem{Horava:1996qa}
P.~Horava and E.~Witten,
``Heterotic and type I string dynamics from eleven dimensions,''
Nucl.\ Phys.\ B {\bf 460}, 506 (1996)
[hep-th/9510209].


\bibitem{Guth:1981zm}
A.~H.~Guth,
``The Inflationary Universe: A Possible Solution To The Horizon And Flatness Problems,''
Phys.\ Rev.\ D {\bf 23}, 347 (1981).


\bibitem{Kallosh:2001ai}
R.~Kallosh, L.~Kofman and A.~Linde,
``Pyrotechnic universe,''
hep-th/0104073.


\bibitem{Khoury:2001iy}
J.~Khoury, B.~A.~Ovrut, P.~J.~Steinhardt and N.~Turok,
``A brief comment on 'The pyrotechnic universe',''
hep-th/0105212.


\bibitem{Donagi:2001fs}
R.~Y.~Donagi, J.~Khoury, B.~A.~Ovrut, P.~J.~Steinhardt and N.~Turok,
``Visible branes with negative tension in heterotic M-theory,''
hep-th/0105199.


\bibitem{Kallosh:2001du}
R.~Kallosh, L.~Kofman, A.~Linde and A.~Tseytlin,
``BPS branes in cosmology,''
hep-th/0106241.

\bibitem{Enqvist:2001}
K.~Enqvist, E.~Keski-Vakkuri and S.~Rasanen, 
``Hubble Law and Brane Matter after Ekpyrosis,''
hep-th/0106282.

\bibitem{Turok:1996wa}
N.~Turok,
``A Causal Source which Mimics Inflation,''
Phys.\ Rev.\ Lett.\  {\bf 77}, 4138 (1996)
[astro-ph/9607109].


\bibitem{Avelino:2000iy}
P.~P.~Avelino and C.~J.~Martins,
``Primordial adiabatic fluctuations from cosmic defects,''
Phys.\ Rev.\ Lett.\  {\bf 85}, 1370 (2000)
[astro-ph/0002413].

\bibitem{Copeland:1997ug}
E.~J.~Copeland, R.~Easther and D.~Wands,
``Vacuum fluctuations in axion-dilaton cosmologies,''
Phys.\ Rev.\ D {\bf 56}, 874 (1997)
[hep-th/9701082].


\bibitem{Guth:1982ec}
A.~H.~Guth and S.~Y.~Pi,
``Fluctuations in the New Inflationary Universe,''
Phys.\ Rev.\ Lett.\  {\bf 49}, 1110 (1982).


\bibitem{Lyth:2001pf}
D.~H.~Lyth,
``The primordial curvature perturbation in the ekpyrotic universe,''
hep-ph/0106153.


\bibitem{Gasperini:1993em}
M.~Gasperini and G.~Veneziano,
``Pre - big bang in string cosmology,''
Astropart.\ Phys.\  {\bf 1}, 317 (1993)
[hep-th/9211021].


\bibitem{Brustein:1995kn}
R.~Brustein, M.~Gasperini, M.~Giovannini, V.~F.~Mukhanov and G.~Veneziano,
``Metric perturbations in dilaton driven inflation,''
Phys.\ Rev.\ D {\bf 51}, 6744 (1995)
[hep-th/9501066].

\bibitem{Hwang:1991an}
J.~Hwang and E.~T.~Vishniac, "Gauge-Invariant Joining Conditions for
Cosmological Perturbations,"
Astrophs.\ J.\ {\bf 382}, 363 (1991).

\bibitem{Deruelle:1995kd}
N.~Deruelle and V.~F.~Mukhanov,
``On matching conditions for cosmological perturbations,''
Phys.\ Rev.\ D {\bf 52}, 5549 (1995)
[gr-qc/9503050].

\bibitem{Martin:1998zd}
J.~Martin and D.~J.~Schwarz,
``The influence of cosmological transitions on the evolution of density  
perturbations,''
Phys.\ Rev.\ D {\bf 57}, 3302 (1998)
[gr-qc/9704049].

\bibitem{Axenides:1983hj}
M.~Axenides, R.~Brandenberger and M.~Turner,
``Development of Axion Perturbations in an Axion Dominated Universe,''
Phys.\ Lett.\ B {\bf 126}, 178 (1983).

\bibitem{Linde:1984ti}
A.~D.~Linde,
``Generation Of Isothermal Density Perturbations In The Inflationary 
Universe,''
JETP Lett.\  {\bf 40}, 1333 (1984)
[Pisma Zh.\ Eksp.\ Teor.\ Fiz.\  {\bf 40}, 496 (1984)].

\bibitem{Easson:2000mj}
D.~A.~Easson,
``The interface of cosmology with string and M(illennium) theory,''
hep-th/0003086.


\bibitem{Brandenberger:2001ph}
R.~H.~Brandenberger,
``The promise of string cosmology,''
hep-th/0103156.

\bibitem{Brandenberger:1989aj}
R.~Brandenberger and C.~Vafa,
``Superstrings in the Early Universe,''
Nucl.\ Phys.\ B {\bf 316}, 391 (1989).

\bibitem{lidsey}
J.~E.~Lidsey, D.~Wands and E.~J.~Copeland, 
``Superstring Cosmology,''
Phys.\ Rept.\ {\bf 337}, 343 (2000).
 
\bibitem{Alexander:2000xv}
S.~Alexander, R.~Brandenberger and D.~Easson,
``Brane gases in the early universe,''
Phys.\ Rev.\ D {\bf 62}, 103509 (2000)
[hep-th/0005212].


\bibitem{Buonanno:1999bi}
A.~Buonanno, T.~Damour and G.~Veneziano,
``Pre-big bang bubbles from the gravitational instability of generic  string vacua,''
Nucl.\ Phys.\ B {\bf 543}, 275 (1999)
[hep-th/9806230].


\bibitem{Lukas:1998fg}
A.~Lukas, B.~A.~Ovrut and D.~Waldram,
``On the four-dimensional effective action of strongly coupled heterotic  string theory,''
Nucl.\ Phys.\ B {\bf 532}, 43 (1998)
[hep-th/9710208];\\
A.~Lukas, B.~A.~Ovrut, K.~S.~Stelle and D.~Waldram,
``The universe as a domain wall,''
Phys.\ Rev.\ D {\bf 59}, 086001 (1999)
[hep-th/9803235].


\bibitem{Moore:2000fs}
G.~Moore, G.~Peradze and N.~Saulina,
``Instabilities in heterotic M-theory induced by open membrane  instantons,''
hep-th/0012104.


\bibitem{Wang:1997cw}
L.~Wang, V.~F.~Mukhanov and P.~J.~Steinhardt,
``On the problem of predicting inflationary perturbations,''
Phys.\ Lett.\ B {\bf 414}, 18 (1997)
[astro-ph/9709032].

\bibitem{Bardeen:1980kt}
J.~M.~Bardeen,
``Gauge Invariant Cosmological Perturbations,''
Phys.\ Rev.\ D {\bf 22}, 1882 (1980).

\bibitem{Mukhanov:1992me}
V.~F.~Mukhanov, H.~A.~Feldman and R.~H.~Brandenberger,
``Theory of cosmological perturbations,''
Phys.\ Rept.\  {\bf 215}, 203 (1992).

\bibitem{Bardeen:1983qw}
J.~M.~Bardeen, P.~J.~Steinhardt and M.~S.~Turner,
``Spontaneous Creation of Almost Scale - Free Density Perturbations in an Inflationary Universe,''
Phys.\ Rev.\ D {\bf 28}, 679 (1983).

\bibitem{Brandenberger:1984tg}
R.~Brandenberger and R.~Kahn,
``Cosmological Perturbations in Inflationary Universe Models,''
Phys.\ Rev.\ D {\bf 29}, 2172 (1984).

\bibitem{Lyth:1985gv}
D.~H.~Lyth,
``Large Scale Energy Density Perturbations and Inflation,''
Phys.\ Rev.\ D {\bf 31}, 1792 (1985).

\bibitem{Bassett:1999wg}
B.~A.~Bassett, D.~I.~Kaiser and R.~Maartens,
``General relativistic preheating after inflation,''
Phys.\ Lett.\ B {\bf 455}, 84 (1999)
[hep-ph/9808404].


\bibitem{Finelli:1999bu}
F.~Finelli and R.~Brandenberger,
``Parametric amplification of gravitational fluctuations during  reheating,''
Phys.\ Rev.\ Lett.\  {\bf 82}, 1362 (1999)
[hep-ph/9809490].


\bibitem{Parry:1999pn}
M.~Parry and R.~Easther,
``Preheating and the Einstein field equations,''
Phys.\ Rev.\ D {\bf 59}, 061301 (1999)
[hep-ph/9809574].


\bibitem{Bassett:2000cg}
B.~A.~Bassett and F.~Viniegra,
``Massless metric preheating,''
Phys.\ Rev.\ D {\bf 62}, 043507 (2000)
[hep-ph/9909353].


\bibitem{Finelli:2000ya}
F.~Finelli and R.~Brandenberger,
``Parametric amplification of metric fluctuations during reheating in two  field models,''
Phys.\ Rev.\ D {\bf 62}, 083502 (2000)
[hep-ph/0003172].

\bibitem{Mukhanov:1988jd}
V.~F.~Mukhanov,
``Quantum Theory Of Gauge Invariant Cosmological Perturbations,''
Sov.\ Phys.\ JETP {\bf 67}, 1297 (1988)
[Zh.\ Eksp.\ Teor.\ Fiz.\  {\bf 94N7}, 1 (1988)].

\bibitem{Sasaki:1986hm}
M.~Sasaki,
``Large Scale Quantum Fluctuations In The Inflationary Universe,''
Prog.\ Theor.\ Phys.\  {\bf 76}, 1036 (1986).

\bibitem{stewlyth:1992} D.~H.~Lyth and E.~D.~Stewart, "The Curvature
Perturbation in Power Law (e.g. Extended) Inflation", 
Phys.\ Lett.\ B {\bf 274}, 168 (1992).

\bibitem{abrasteg} M. Abramowicz and I. A. Stegun, {\em Handbook of
Mathematical Functions} (Dover, New York, 1972).

\bibitem{Kost4} J. Khoury, B. Ovrut, P. Steinhardt and N. Turok,
``Gravitational backreaction and density perturbations in
ekpyrotic models'', to be submitted.

\bibitem{Lyth2} D. Lyth,
``The curvature perturbation in the ekpyrotic and pre-big-bang
Universes'', to be submitted.

\bibitem{vandeBruck:2000ju}
C.~van de Bruck, M.~Dorca, R.~H.~Brandenberger and A.~Lukas,
``Cosmological perturbations in brane-world theories: Formalism,''
Phys.\ Rev.\ D {\bf 62}, 123515 (2000)
[hep-th/0005032].

\bibitem{Maartens:2000fg}
R.~Maartens,
``Cosmological dynamics on the brane,''
Phys.\ Rev.\ D {\bf 62}, 084023 (2000)
[hep-th/0004166].

\bibitem{Langlois:2000ia}
D.~Langlois,
``Brane cosmological perturbations,''
Phys.\ Rev.\ D {\bf 62}, 126012 (2000)
[hep-th/0005025].

\bibitem{Kodama:2000fa}
H.~Kodama, A.~Ishibashi and O.~Seto,
``Brane world cosmology: Gauge-invariant formalism for perturbation,''
Phys.\ Rev.\ D {\bf 62}, 064022 (2000)
[hep-th/0004160].

\bibitem{Kost5} 
J.~Khoury, B.~A.~Ovrut, N. Seiberg, P.~J.~Steinhardt and N.~Turok,
``From Big Crunch To Big Bang'', hep-th/0108187.

\end{thebibliography}
\end{document}